\documentclass[reprint, superscriptaddress, twocolumn, amsmath,amssymb,aps, pra,]{revtex4-2}

\usepackage[dvipsnames]{xcolor}
\usepackage{graphicx}% Include figure files
\usepackage{dcolumn}% Align table columns on decimal point
\usepackage{bm}% bold math
\usepackage{hyperref}% add hypertext capabilities
\usepackage{placeins}
\usepackage{caption}
\usepackage[mathlines]{lineno}
\usepackage{siunitx}
\usepackage{braket}

\DeclareSIUnit\gauss{G}

\begin{document}

\preprint{APS/123-QED}

\title{Low-field Feshbach resonances and three-body losses\\
in a fermionic quantum gas of $^{161}$Dy}

\author{E. Soave}
\affiliation{Institut f{\"u}r Experimentalphysik, Universit{\"a}t Innsbruck, Austria} 

\author{V. Corre}
\altaffiliation[Present address: ]
{Safran Reosc, Sain-Pierre-du-Perray 91280, France}
\affiliation{Institut f{\"u}r Experimentalphysik, Universit{\"a}t Innsbruck, Austria} 

\author{C. Ravensbergen}
\altaffiliation[Present address: ]
{ColdQuanta,
Oxford Centre for Innovation, OX1 1BY, United Kingdom}
\affiliation{Institut f{\"u}r Experimentalphysik, Universit{\"a}t Innsbruck, Austria} 
 
\author{J. H. Han}
\altaffiliation[Present address: ]
{Korea Research Institute of Standards and Science, Daejeon 34113, South Korea}
\affiliation{Institut f{\"u}r Experimentalphysik, Universit{\"a}t Innsbruck, Austria} 

\author{M. Kreyer}
\affiliation{Institut f{\"u}r Experimentalphysik, Universit{\"a}t Innsbruck, Austria} 

\author{E. Kirilov}
\affiliation{Institut f{\"u}r Experimentalphysik, Universit{\"a}t Innsbruck, Austria} 
\affiliation{Institut f{\"u}r Quantenoptik und Quanteninformation (IQOQI), {\"O}sterreichische Akademie der Wissenschaften, Innsbruck, Austria}

\author{R. Grimm}
\affiliation{Institut f{\"u}r Experimentalphysik, Universit{\"a}t Innsbruck, Austria} 
\affiliation{Institut f{\"u}r Quantenoptik und Quanteninformation (IQOQI), {\"O}sterreichische Akademie der Wissenschaften, Innsbruck, Austria}

\date{\today}

\begin{abstract}
We report on high-resolution Feshbach spectroscopy on a degenerate, spin-polarized Fermi gas of $^{161}$Dy atoms, measuring three-body recombination losses at low magnetic field. For field strengths up to 1\,G, we identify as much as 44 resonance features and observe plateaus of very low losses. For four selected typical resonances, we study the dependence of the three-body recombination rate coefficient on the magnetic resonance detuning and on the temperature. We observe a strong suppression of losses with decreasing temperature already for small detunings from resonance. The characterization of complex behavior of three-body losses of fermionic $^{161}$Dy is important for future applications of this peculiar species in research on atomic quantum gases.
\end{abstract}
\maketitle
\section{\label{Introduction}Introduction}
Over the past decade, the exotic interactions of submerged-shell lanthanide atoms have tremendously boosted experimental research on ultracold quantum gases \cite{Chomaz2022dpa}. Exciting properties of such atoms result from long-range anisotropic interactions in combination with tunability of the contact interaction via magnetically controlled Feshbach resonances~\cite{Chin2010fri}. Prominent examples for novel states of matter created in the laboratory are quantum ferrofluids of Dy \cite{Kadau2016otr} and supersolids realized with both Dy and Er \cite{Tanzi2019ooa, Bottcher2019tsp, Chomaz2019lla}. Progress has also been made with quantum-gas mixtures of different lanthanide atoms (Dy-Er) \cite{Trautmann2018dqm,Politi2022iii} and mixtures of lanthanide and alkali-metal atoms (Dy-K) \cite{Ravensbergen2018poa, Ravensbergen2020rif}, with a wide potential for future experiments on exotic states of quantum matter.

For interaction control, magnetic lanthanide atoms offer a rich spectrum of Feshbach resonances \cite{Frisch2014qci,Baumann2014ool,Burdick2016lls,Maier2015eoc}, much denser as compared to alkali-metal atoms. This experimentally well-established fact is a consequence of anisotropy stemming both from the strong magnetic dipole-dipole interaction and from the van-der-Waals interaction for electronic ground states with non-zero orbital angular momentum \cite{Petrov2012aif,Kotochigova2014cib}. The anisotropic interaction leads to a strong mixing of different partial waves. If hyperfine structure is present, such as for the fermionic isotopes $^{161}$Dy and $^{167}$Er, the Feshbach spectrum is even more complex, and the blessing of tunability may turn into a curse of omnipresent three-body recombination losses.

The experiments performed with $^{161}$Dy in our laboratory are motivated by the prospect to realize novel superfluid states in mass-imbalanced fermion mixtures \cite{Gubbels2013ifg, Wang2017eeo, Pini2021bmf}. In a Fermi-Fermi mixture of $^{161}$Dy and $^{40}$K atoms, we have recently demonstrated hydrodynamic behavior as a manifestation of strong interactions, realized on top of an interspecies Feshbach resonance \cite{Ravensbergen2020rif}. Further experiments are in progress on the formation of bosonic Feshbach molecules, paired fermionic many-body states, and collective behavior of the strongly interacting mixture. In all these experiments, the Dy-Dy intraspecies Feshbach resonances represent a complication and appropriate strategies have to be developed to minimize unwanted effects, such as three-body losses and heating. 
%This, in turn, requires detailed knowledge on the general behavior of three-body recombination losses in the ultradense Feshbach spectrum of this fermionic species. %$^{161}$Dy.

Feshbach resonances in spin-polarized fermionic quantum gases result 
from scattering in odd partial waves. Accordingly, $p$-wave resonances have been observed in early experimental work \cite{Regal2003tpw,Zhang2004pwf,Schunck2005fri} and studied theoretically \cite{Suno2003rot,Chevy2005rsp}. More recent experiments \cite{Yoshida2018SLF,Waseem2018ULB} have provided deeper insights into the scaling laws and universal properties of three-body recombination losses near $p$-wave resonances. Our present situation of $^{161}$Dy, however, is more complex because of the strong coupling between different partial waves and the possible interaction between different closely spaced or overlapping resonances, which makes a theoretical description very challenging. Experiments are needed to find out to what extent our resonances in $^{161}$Dy behave in a similar way.

In this article, we report on the experimental investigation of the ultradense Feshbach spectrum of $^{161}$Dy at low magnetic field strength (up to about 1\,G) with high resolution ($\sim$1\,mG). To minimize the effect of finite collision energies, i.e.\ broadening effects and the influence of higher partial waves, we work in the deeply quantum-degenerate regime at rather low values of the Fermi energy down to a few $100$\,nK. In Sec.~\ref{SecSpectrum}, we present the Feshbach loss spectrum, exhibiting nearly 50 loss features in a 1\,G wide range. We also identify plateaus of very low losses, which can be used for efficient evaporative cooling. In Sec.~\ref{paragra:K3studyOfSelectedResonances}, we then present case studies on four typical resonances, where we report on the dependence of the three-body rate coefficient on the magnetic detuning and the temperature of the sample. Our measurements show that even very small detunings from resonance of a few mG are sufficient to enter a regime where losses are strongly suppressed with decreasing temperature.

\section{\label{SecSpectrum}Low-field Feshbach Spectrum}
\subsection{\label{Subsec:SamplePreparation}Sample preparation}
All our experiments begin with the production of a degenerate Fermi gas of $^{161}$Dy atoms. We follow the procedures described in detail in Ref.\,\cite{Ravensbergen2018poa}: After capturing the atoms in a magneto-optical trap (MOT) operated at the 626-nm intercombination line\,\cite{MaierNLM2014}, the sample is transferred into a crossed-beam optical dipole trap (ODT), which uses near-infrared light at a wavelength of \SI{1064}{\nano\meter}. Here forced evaporative cooling is performed by ramping down the trapping potential. Under optimized conditions, we obtain a sample of up to $N=\SI{1.5e5}{}$ atoms in a nearly harmonic trap (geometrically averaged trap frequency $\bar{\omega}=2\pi\times\SI{150}{\hertz}$) at a temperature of $T=\SI{80}{\nano\kelvin}$. With a Fermi temperature of $T_F=\hbar\bar{\omega}(6N)^{1/3}/k_B=\SI{695}{\nano\kelvin}$, this corresponds to deeply degenerate conditions with $T/T_F=0.12$ and a peak number density of $\hat{n}=\SI[per-mode=symbol]{1.6e14}{\cm\tothe{-3}}$ in the center of the trap. Our sets of measurements are taken over typically many hours (sometimes even a few days), where long-term drifts may reduce the maximum atom number provided by roughly a factor of two. In a last preparation stage, the ODT is modified by replacing one of the laser beams (horizontally propagating) with a beam of larger waist. This modification provides us with more flexibility to vary the trap frequency $\bar{\omega}$ and, in particular, it allows us to realize very shallow traps to work at lower atomic number densities. For each experiment, the trap is chosen in a way to avoid residual evaporation. The particular conditions for each set of measurements are listed in App.\,\ref{App:Initialconditions}.
\par The cloud is fully spin polarized in the lowest hyperfine Zeeman sub-level $\ket{F,m_F}=\ket{21/2,-21/2}$ as a result of optical pumping during the MOT stage \cite{Dreon2017DOA} and subsequent rapid dipolar relaxation of residual population in higher spin states in the ODT \cite{Lu2012QDD}. 
For the fully spin-polarized sample, inelastic two-body losses are suppressed already at very low magnetic field values. The minimization of three-body losses, essential for efficient evaporative cooling, depends very sensitively on the particular magnetic field applied. Our evaporation sequence is performed at a magnetic bias field of \SI{230}{\milli\gauss}, which we found to work slightly better than at \SI{430}{\milli\gauss}, as applied in Ref.\,\cite{Ravensbergen2018poa}.
\subsection{Loss Scan\label{Sec:LossScan}}
\begin{figure*}
\centering
\includegraphics[trim=80 0 80 0,clip,width=2\columnwidth]{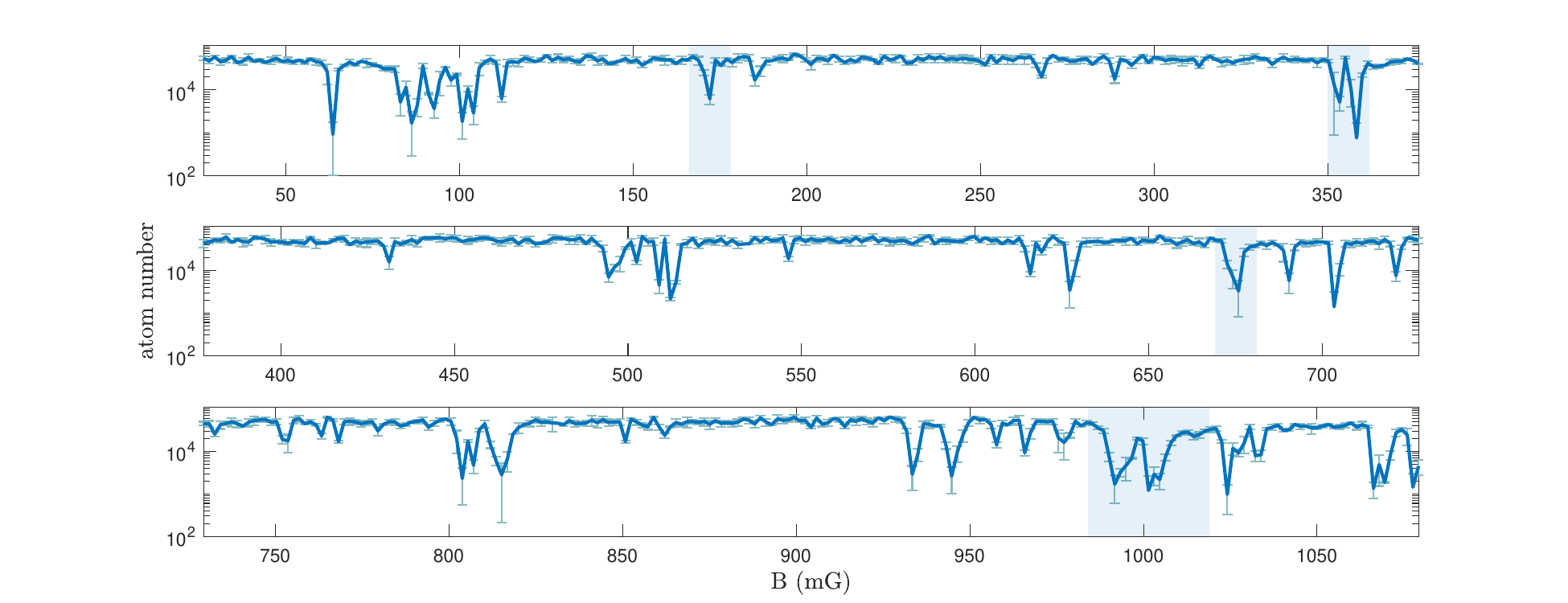}
%\captionsetup{justification=raggedright,singlelinecheck=false}
\caption[.]{Low-field Feshbach spectrum of a degenerate sample of spin-polarized $^{161}$Dy. The magnetic field is varied in steps of \SI{1.6}{\milli\gauss}. The error bars show the sample standard deviation of three individual measurements at the same magnetic field. We observe about 44 loss features, which we attribute to Feshbach resonances. The shaded areas correspond to the four resonances that are investigated in detail in Sec.\,\ref{paragra:K3studyOfSelectedResonances}.}\label{fig:LowFieldRegionFBScan}
\end{figure*}
We study the low-field Feshbach spectrum by measuring atom losses for a variable magnetic field strength \cite{Chin2010fri} in the range between 0 and \SI{1}{\gauss}. After preparation of the sample in a very shallow ODT (for experimental parameters see App.\,\ref{App:Initialconditions}), we ramp the magnetic field from the evaporation field to the variable target one in \SI{20}{\milli\second}. The low trap frequency of $\bar{\omega}=2\pi\times\SI{100}{\hertz}$ is chosen to minimize losses induced by the magnetic field ramp. We hold the cloud for \SI{7}{\second}, and then release it from the ODT. An absorption image is taken after a time of flight of \SI{10}{\milli\second}.
\par The magnetic-field stability is essential for resolving narrow loss features. Using radio-frequency spectroscopy\footnote{We investigate the magnetic-field stability by performing radio-frequency spectroscopy on $^{40}$K. The possibility to work with potassium in the same setup follows from the fact that our apparatus is designed for mixture experiments \cite{Ravensbergen2018poa,Ravensbergen2020rif}.} we identified a 50-Hz ripple in the ambient magnetic field as the main source of noise, with a peak-to-peak value of \SI{1.7}{\milli\gauss}. Other noise sources, such as noise in the current of our coils, stay well below an estimated rms level of \SI{1}{\milli\gauss}.
\par In Fig.~\ref{fig:LowFieldRegionFBScan} we plot the remaining atom number as a function of the magnetic field. We count $\simeq44$ loss features, which we assign to Feshbach resonances. On resonance the three-body recombination rate is greatly enhanced and leads to more than a factor of 10 reduction in atom number. At these positions we also observe substantial heating (not shown). The resonances seem to mostly gather in groups, with flat, typically tens of \SI{}{\milli\gauss} wide, plateaus between them. Within these plateaus, losses are rather weak and stay within a few percent even for the long hold time of \SI{7}{\second} applied.
\par The recorded Feshbach spectrum resembles previous observations in submerged-shell lanthanide atoms (Er \cite{Frisch2014qci, Maier2015eoc}, Dy \cite{Baumann2014ool, Burdick2016lls}, Tm \cite{RTC2019Khlebnikov}), which are known to exhibit a dense and very complex resonance spectrum. In the cases of the fermionic isotopes $^{161}$Dy and $^{167}$Er, where hyperfine structure is present, the resonance density is extremely high. While for $^{167}$Er a density of about 25 resonances per gauss has been reported in the range between 0 and \SI{4.5}{\gauss} \cite{Frisch2014qci}, previous work on $^{161}$Dy has revealed between about 10 resonances per gauss in a range between 0 and \SI{6}{\gauss} \cite{Baumann2014ool} and up to about 100 resonances in a 250-mG wide range near \SI{34}{\gauss} \cite{Burdick2016lls}. With our 44 resonances in a range between 0 and \SI{1}{\gauss}, we apparently resolve more resonances than in Ref.\,\cite{Baumann2014ool}, which we attribute to our higher magnetic field resolution. We believe that a further improved magnetic field stability to well below \SI{2}{\milli\gauss} would reveal even more resonances and a substructure of some of our observed features. The complex spectrum of resonances may be further analyzed using statistical methods \cite{Maier2015eoc, RTC2019Khlebnikov}, which is beyond the scope of the present work.

\section{\label{paragra:K3studyOfSelectedResonances}Case Studies of Selected Resonances}
We now perform a systematic investigation of the $K_3$ coefficient as a function of the magnetic-field and the temperature for selected resonances. In Sec.\,\ref{SubSec:K3calculation} we first show how, from atom number decay measurements, we obtain the value of the three-body recombination coefficient $K_3$. In Secs.\,\ref{MFDependence} and \ref{TemperatureDependence} we investigate the dependence of $K_3$ on the magnetic field and the initial temperature, respectively.

\subsection{\label{SubSec:K3calculation}Three-body decay curves and loss-rate coefficients}
In the absence of two-body losses, the evolution of the number of trapped atoms $N(t)$ can be modeled based on the differential equation
\begin{equation}\label{eqAtomNumberRateEquation}  
    \dot N(t) = -\Gamma_{v} N(t)-3K_3\int d^3r\,n^3(\bold{r},t),
\end{equation}
where $\Gamma_v$ is the one-body loss rate from collisions with rest-gas particles, and $n(\bold{r},t)$ represents the number density distribution of the cloud. The quantity $K_3$ denotes the three-body event rate coefficient, which for a single atomic species is related to the commonly used three-body loss rate coefficient by $L_3=3K_3$. Note that, according to our phenomenological definition, the $K_3$ coefficient represents a thermal average over the distribution of collision energies in the sample, and does not represent the coefficient for a specific collision energy as used in theoretical work \cite{TLFEsry2001}. For our experiments we estimate a rest-gas limited lifetime as long as $1/\Gamma_v \simeq \SI{60}{\second}$. Given such a low one-body loss rate, $\Gamma_v$ can be neglected in the analysis of near-resonance decay curves, while it is relevant for cases on the long-lived plateaus.
\begin{figure}[t]
\centering
\includegraphics[trim=2 0 2 0,clip,width=1\columnwidth]{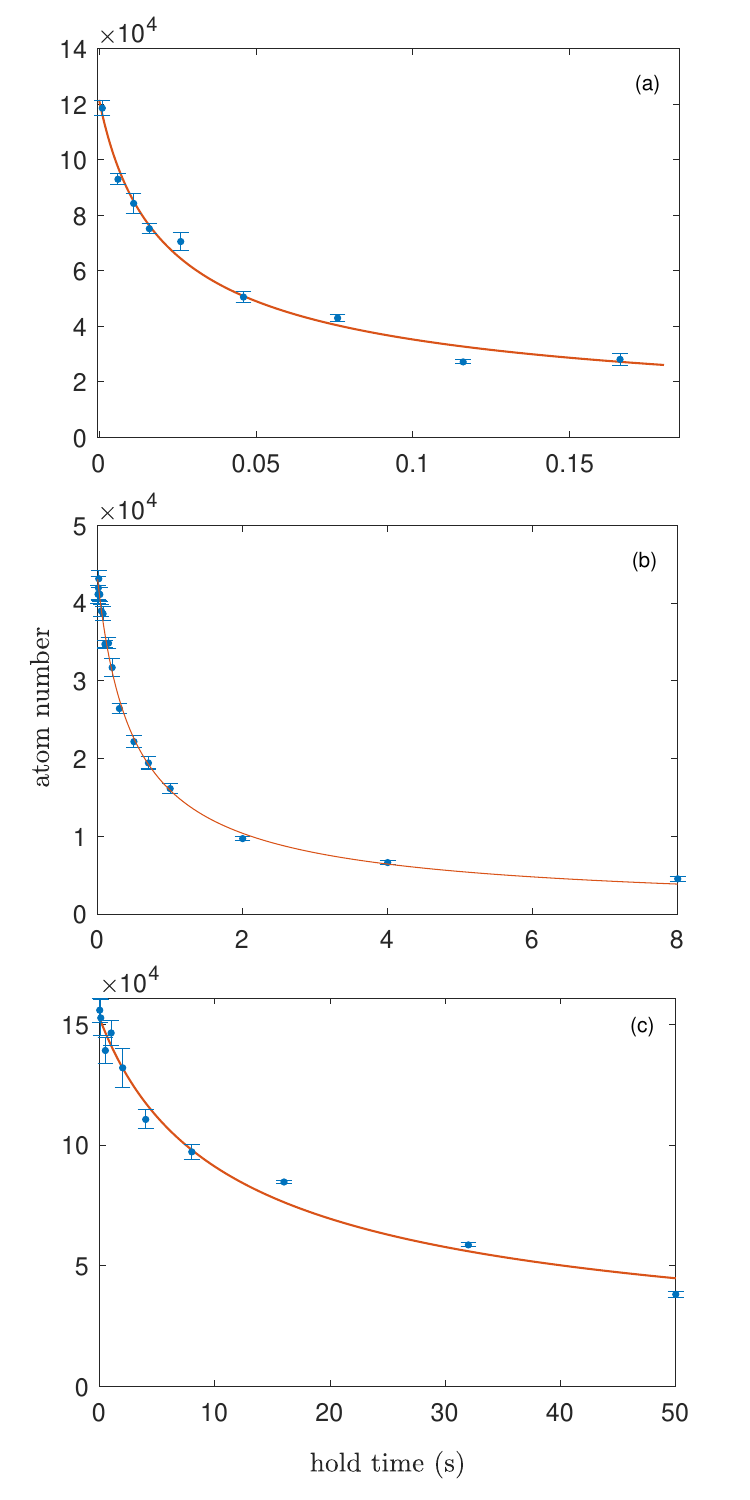}
%\captionsetup{justification=raggedright,singlelinecheck=false}
\caption{\label{FigExampleMeasurement}Typical decay curves. The measurements have been performed (a) on resonance at $B=\SI{678}{\milli\gauss}$, (b) for small detuning at $B=\SI{989}{\milli\gauss}$, and (c) on a minimum-loss plateau at $B=\SI{225}{\milli\gauss}$. The solid lines show fits by the heuristic model introduced in App.\,\ref{App:K3derivation}.}
\end{figure}
\par In Fig.~\ref{FigExampleMeasurement} we show three typical decay curves, on resonance (a), near a resonance (b) and far away from any resonance (c). The sample is held in the ODT at a fixed magnetic field. After a variable hold time the cloud is released and the number of remaining atoms is measured by time-of-flight imaging. 
To analyze the decay curves and to extract values for $K_3$, we apply a heuristic model (for details see App.\,\ref{App:K3derivation}) to quantify the initial slope $\dot N(0)$. From the initial decay rate $1/\tau=-\dot N(0)/N(0)$ and knowledge of the experimental parameters at $t=0$, we then calculate the resulting values for $K_3$. This approach, which focuses on the initial decay, avoids complications by the heating of the sample during the decay. Depending on the experimental conditions under consideration, decay times can vary from a few ms to many seconds.
%, as can be inferred from the comparison of Fig. 2.a (on resonance, $\tau=0.50(9)\,$s) and Fig. 2.b (far detuned, $\tau=12(3)\,$s). 
As an example, the measurement reported in Fig.\,\ref{FigExampleMeasurement}(a) was carried out under typical experimental conditions (see App.\,\ref{App:Initialconditions}) very close to the center of the 679-mG resonance, with initially about \SI{1.2e5}{} atoms. Our fit yields an initial decay time $\tau=\SI{22(6)}{\milli\second}$, and for the three-body rate coefficient we obtain $K_3=\SI[per-mode=symbol]{4(1)e-26}{\cm\tothe{6}\per\second}$.
The same measurement, performed few mG detuned from the resonance at \SI{995}{\milli\gauss} and reported in Fig.\,\ref{FigExampleMeasurement}(b), already shows a significant longer decay time ($\tau=\SI{0.50(9)}{\second}$). We calculate a three-body recombination coefficient value $K_3=\SI[per-mode=symbol]{7(2)e-28}{\cm\tothe{6}\per\second}$, two orders of magnitude lower than on resonance. 
\par The measurement in Fig.\,\ref{FigExampleMeasurement}(c) is carried out at a magnetic field of \SI{225}{\milli\gauss}, on a minimum-loss plateau, and reveals a very long lifetime. To observe the effect of three-body losses we worked in a tightly compressed trap with $\bar{\omega}=2\pi\times\SI{380}{\hertz}$, leading to a peak-density of $\hat{n}_0\approx\SI[per-mode=symbol]{6e14}{\cm\tothe{-3}}$, which is exceptionally high for a degenerate Fermi gas. We measure an initial decay time $\tau=\SI{12(3)}{\second}$, from which a value $K_3= \SI[per-mode=symbol]{5(3)e-32}{\cm\tothe{6}\per\second}$ is obtained. This value is extraordinary low, which is highlighted by comparison with $^{87}$Rb as a widely used bosonic species, where the $K_3$ coefficient has been measured to be of the order of $10^{-29}\SI[per-mode=symbol]{}{\cm\tothe{6}\per\second}$ \cite{CCA1997Burt, Soding1999TBD}. Such an extremely weak three-body decay, together with the sizeable elastic scattering cross section from dipolar collisions \cite{Bohn2009QUD}, explains why Fermi gases of submerged-shell lanthanide atoms facilitate highly efficient evaporative cooling \cite{Aikawa2014RFD, Ravensbergen2018poa}.

\subsection{\label{MFDependence}Dependence on magnetic-field detuning}
\begin{figure*}[t]
\centering
\includegraphics[width=2\columnwidth]{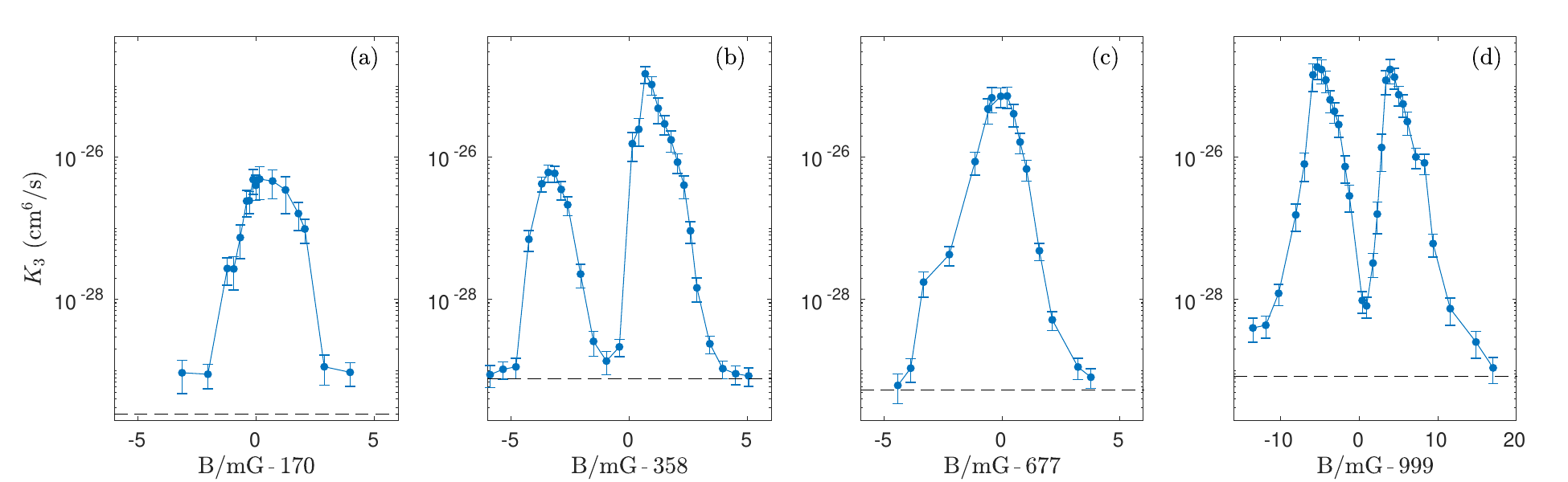}
\caption{$K_3$ coefficient as a function of the magnetic-field, measured for the four selected resonances.
Note that in (a)-(c) the full range covers 12\,mG, whereas in (d) it is three times wider. Each data point is obtained from an individual decay curve as discussed in Sec.\,\ref{SubSec:K3calculation}. The dashed lines indicate the lower limit to the measurable $K_3$ value, imposed by one-body losses.}
\label{FigThreeResonancesComparison} 
\end{figure*}
In this Section, we discuss selected loss features as typical examples for the many resonances observed in our Feshbach spectrum. We focus on three resonant features that lead to relatively strong losses in the measured Feshbach spectrum (near \SI{358}{\milli\gauss}, \SI{677}{\milli\gauss} and \SI{999}{\milli\gauss}, see loss scan in Fig.\,\ref{fig:LowFieldRegionFBScan}). For reference, we also investigate a weaker loss resonance (near \SI{170}{\milli\gauss}), which appears to be well isolated from other resonances. We consider their line shapes and widths by presenting measurements on the $K_3$ values as a function of the magnetic detuning from resonance. Here we work in the deeply degenerate regime, with $T/T_F\simeq0.2$ at a low $T_F\simeq\SI{400}{\nano\kelvin}$ (for details see App.~\ref{App:Initialconditions}), which minimizes line broadening stemming from the finite kinetic energy \cite{Maier2015eoc,Green2020FRP}.
Our results are displayed in Fig.~\ref{FigThreeResonancesComparison}(a-d). The measured values for $K_3$ vary over more than four orders of magnitude. Maximum values are found to exceed \SI[per-mode=symbol]{e-25}{\cm\tothe{6}\per\second}. The presence of weak one-body losses (see Sec.~\ref{SubSec:K3calculation}) imposes a lower limit for the measurable $K_3$ value, which, for this particular trapping conditions, is in the range of a few \SI[per-mode=symbol]{e-30}{\cm\tothe{6}\per\second}. This lower limit is indicated by the dashed horizontal lines in Fig.~\ref{FigThreeResonancesComparison}. 
\par Figure~\ref{FigThreeResonancesComparison}(a) shows the resonance near \SI{170}{\milli\gauss}, which is the weakest of the four selected features. We observe a full width of about \SI{3.5}{\milli\gauss} \footnote{We define the width as the full magnetic field range where the $K_3$ value exceeds the geometric average between its maximum and minimum. This corresponds to the full width at half maximum on a logarithmic scale.}. The line shape is essentially symmetric, which may first appear surprising in view of the expected asymmetric line shapes of Feshbach resonances in higher partial waves, which usually show a sharp edge on the lower side (marking the resonance position) along with a tail on the upper side \cite{Green2020FRP,Ticknor2004MSF,Nakasuji2013EDP,Jiaming2018TBR}.
We assume that the shape of the weak feature is dominated by the magnetic-field fluctuations in our experimental setup (see Sec.\,\ref{Sec:LossScan}), which may affect the observed loss features in a range of a few mG. The fluctuations will smear out any narrower feature and mask the true resonance line shape (see discussion on broadening effects in App.\,\ref{App:Broadening}). This interpretation is supported by the fact that we never observe any narrower feature. We therefore believe that the observed behavior of narrower resonances, such as the 170-mG feature, is dominated by magnetic-field fluctuations.
\par In Fig.~\ref{FigThreeResonancesComparison}(b) we show a double feature of two resonances, separated by about \SI{4}{\milli\gauss}. While the weaker feature near \SI{355}{\milli\gauss} closely resembles the one in Fig. 2(a), the stronger feature near \SI{359}{\milli\gauss} shows a peak value for $K_3$ exceeding \SI[per-mode=symbol]{e-25}{\cm\tothe{6}\per\second}, which is an order of magnitude higher. The stronger feature also shows indications of the tail expected on the upper side for such resonances. The resonance appears to be wide enough that its true structure is not fully masked by the magnetic-field fluctuations.
Figure~\ref{FigThreeResonancesComparison}(c) shows a feature near \SI{677}{\milli\gauss}, which in the Feshbach scan in Fig.~\ref{fig:LowFieldRegionFBScan}, appeared to be a single, relatively strong resonance. A closer investigation, however, reveals a shoulder on the lower side, which is likely to be caused by another weak overlapping resonance. On the upper side, the $K_3$ coefficient falls off in a way resembling the expected tail.
Figure~\ref{FigThreeResonancesComparison}(d) finally displays our strongest observed loss feature; note the three times wider magnetic-field range. We see a double feature separated by about \SI{10}{\milli\gauss}. The line shapes of the two resonances correspond to the expectation of a sharper edge on the lower side and a tail on the upper side. Here, at least for these broader features, magnetic-field fluctuations do not have a substantial effect on the line shape.

\subsection{\label{TemperatureDependence}Temperature dependence}
We now turn our attention to the dependence of the $K_3$ coefficient on the temperature of the cloud, for different magnetic detunings from the resonance center. We vary the temperature of the cloud by interrupting the evaporation sequence in a controlled way and by adiabatically varying the final trap frequency. For these measurements, decay is observed in a 160-Hz and a 400-Hz trap, for lower and higher temperatures, respectively. The $K_3$ coefficient is obtained according to Eq.(\ref{eqK3ForDegenerateCloud}) or Eq.(\ref{eqK3ForThermalCloud}), depending on the initial $T/T_F$ of the sample. 
We introduce the effective temperature $\tilde{T}$, such that the mean energy per particle is $3k_B\tilde{T}$. This definition takes into account that, for a degenerate Fermi gas, the relative momentum and thus the collision energy stay finite even at $T=0$. In the limit of a thermal gas, $\tilde{T}=T$ holds, while for a degenerate one we have $\tilde{T}=T\,\mathrm{Li}_4(-\zeta)/\mathrm{Li}_3(-\zeta)$ \cite{demarco2001thesis}, where $\zeta$ is the fugacity and $\mathrm{Li}_i$ is the polylogarithm of order $i$.
\par
\begin{figure}[t]
\centering
\includegraphics[width=1\columnwidth]{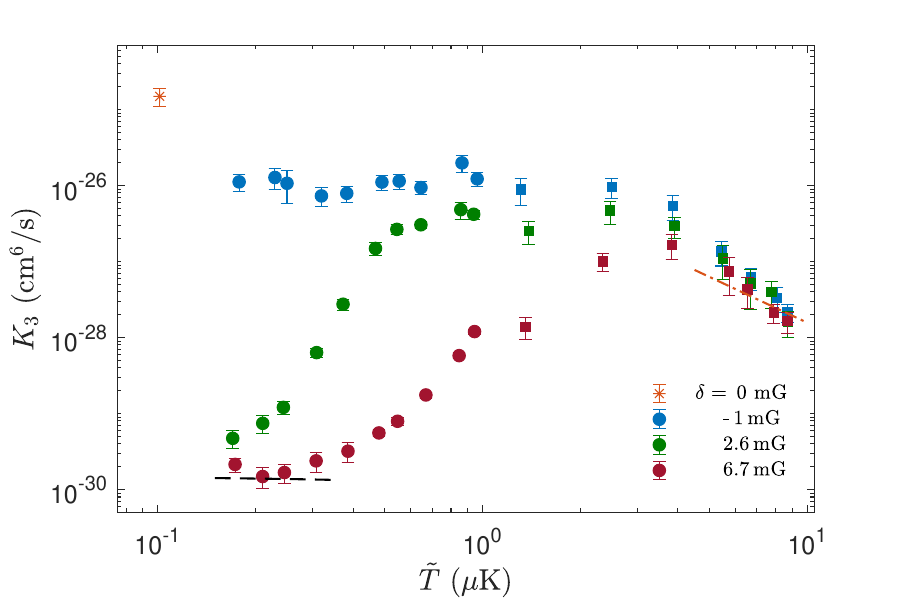}
\caption{Temperature dependence of $K_3$, for various detunings (typical uncertainty 0.2\,mG) relative to the center of the 358-mG resonance. For $\tilde{T}\lesssim\SI{1}{\micro\kelvin}$ the samples have initial $T/T_F<1$. The decay curves have been measured in a $\bar{\omega}/2\pi=$160-Hz (circles) and 400-Hz trap (squares). As a reference, we plot the $K_3$ resonant peak value, corresponding to the maximum in Fig.\,\ref{FigThreeResonancesComparison}(b). The dot-dashed line shown for $T>\SI{5}{\micro\kelvin}$ indicates the $T^{-2}$ dependence, according to Eq.\,(\ref{EqUnitaryRegime}), with $\zeta=0.022$. The dashed line indicates the lower limit to the measurable $K_3$ value, imposed by one-body losses.}
\label{FigTemperatureDependence} 
\end{figure}

 Figure\,\ref{FigTemperatureDependence} reports the behavior of the $K_3$ value as a function of $\tilde{T}$ for three different magnetic detunings relative to the center of the 358-mG resonance.
\par We first discuss the $K_3$ behavior in the high energy regime ($\tilde{T}\gtrsim\SI{3}{\micro\kelvin}$). Here the three curves decrease in a similar manner, showing no dependence on the magnetic field. Such a behavior reflects the unitarity limit for the $K_3$ coefficient, which was predicted and observed in several resonantly interacting systems, fermionic and bosonic ones (e.g. \cite{Rem2013LOT,Fletcher2013SOA,BUFMaier2015,ULDEismann2016,Yoshida2018SLF}). A non-degenerate atomic system enters the unitarity-limited regime when the thermal de Broglie wavelength $\lambda_{\rm{dB}}=\hbar\sqrt{2\pi/(m k_BT)}$ becomes comparable to a characteristic length associated with the resonance at a given magnetic detuning \footnote{The length scale that characterizes the interaction is the scattering length $a$ for an \textit{s}-wave resonance and by $\sqrt{|V_p|k_{\rm{eff}}}$ in the case of \textit{p}-wave Feshbach resonances. Here $V_p$ and $k_{\rm{eff}}$ are the scattering volume and the effective range, respectively.}. We attribute to the competition between those two length scales the fact that the larger the detuning, the higher is the temperature at which the $K_3$ value enters the unitary regime. In this regime the $K_3$ value is expected to scale as $T^{-2}$ \cite{Suno2003rot}: 
\begin{equation}\label{EqUnitaryRegime}
    K_3=\zeta\frac{12\sqrt{3}\pi^2\hbar^5}{m^3(k_BT)^2}.
\end{equation}
The prefactor $\zeta$ relates to the efficiency of three colliding atoms forming a dimer and a free atom, and is believed to be a non-universal (i.e.~species-dependent) quantity. A fit to our data, considering only the points with $\tilde{T}>\SI{5}{\micro\kelvin}$, yields a value $\zeta=0.022(2)$. In Ref.\,\cite{Waseem2018ULB} the authors extracted a value $\zeta=0.09$ for $^6$Li. Those two results are about an order of magnitude below what has been observed for Bose gases, where values of $\zeta=0.9,\,0.3$, and 0.24 have been derived for $^7$Li \cite{Rem2013LOT}, $^{39}$K \cite{Fletcher2013SOA}, and $^{164}$Dy \cite{BUFMaier2015}, respectively.
\par We now discuss the temperature-dependence of $K_3$ far below the unitarity-limited regime. Figure \,\ref{FigTemperatureDependence} demonstrates that even a very small magnetic resonance detuning of a few mG can have a dramatic effect on the low-temperature behavior. The data taken at $\delta = 6.7$\,mG (typical uncertainty 0.2\,mG) show a reduction of $K_3$ from a maximum value of the order of $10^{-27}$\,cm$^6$/s at $\tilde{T} = 3.5\,\mu$K to a minimum of  
about $10^{-30}$\,cm$^6$/s at $\tilde{T} \lesssim 200\,$nK. Note that the minimum value that we can observe is limited by one-body decay (dashed line), so that the true suppression will be even larger. A very similar behavior is observed closer to resonance at $\delta = 2.6$\,mG. Here a maximum $K_3$ value of $\sim 5 \times 10^{-27}$\,cm$^6$/s is found at $\tilde{T}$ of the order of $1\,\mu$K, which is reduced by three orders of magnitude at our lowest temperature $\tilde{T} \approx 200\,$nK. The main effect of the smaller detuning appears to be a shift of the qualitatively similar behavior to lower temperatures.

These observations on the low-temperature behavior can be compared with recent experimental work studying three-body recombination on $p$-wave Feshbach resonances in $^6$Li \cite{Yoshida2018SLF,Waseem2018ULB}. For the limit of very low collision energies $E_{\rm coll}$, a threshold law $K_3 \propto E_{\rm coll}^2$, as originally predicted in Ref.~\cite{Suno2003rot}, has been observed in Ref.~\cite{Yoshida2018SLF} for a thermal ($T/T_F > 1$) Fermi gas, where $K_3 \propto T^2$. This observation of threshold-law behavior required a relatively large resonance detuning. In our case, with rather small detunings, the threshold-law regime would require extremely low collision energies. This regime, however, remains inaccessible in our present experiments because of two limitations: The Fermi energy gives a lower limit to the collision energy ($\tilde{T} = T_F/4$ at $T = 0$), and one-body losses do not allow us to measure $K_3$ values below $\sim 10^{-30}$\,cm$^6$/s. However, beyond the threshold-law regime, we observe the same steep increase with temperature as seen in Ref.~\cite{Yoshida2018SLF} for relatively large magnetic detunings. The breakdown of the threshold law has been interpreted \cite{Yoshida2018SLF} in terms of the effective range of the resonance.

The case very close to resonance (data points for $\delta = -1\,$mG in Fig.~\ref{FigTemperatureDependence}) reveals a different behavior. Here we do not observe any loss suppression with decreasing temperature. The $K_3$ value appears to level off at about $10^{-26}$\,cm$^6$/s. This, however, does not rule out the possibility of loss suppression at values of $\tilde{T}$ that are even lower than what we can realize experimentally in the deeply degenerate situation. The single data point shown for $\delta=0$ at $\tilde{T} \approx 75$\,nK corresponds to the loss maximum in Fig.~\ref{FigThreeResonancesComparison}(b). This measurement highlights that three-body losses can be very strong on top of the resonance, suggesting no suppression at low temperatures. A similar on-resonance behavior has been observed in Refs.~\cite{Waseem2018ULB,Zhang2004pwf}, but in contrast to our work these experiments were limited to the non-degenerate case. 

In the narrow detuning range of $|\delta| \lesssim 1$\,mG, the interpretation of our present results is impeded by the sensitivity of the experiments to magnetic field noise (see Sec.~\ref{Sec:LossScan} and App.~\ref{App:Broadening}). The on-resonance behavior of three-body recombination at ultralow collision energies, which has also been subject to recent theoretical investigations \cite{Schmidt2020TBL}, thus remains a topic for future research.
\section{Conclusion}
In summary, we have carried out Feshbach spectroscopy on an optically trapped spin-polarized degenerate Fermi gas of $^{161}$Dy atoms by measuring three-body recombination losses. We have focused on the range of low magnetic fields up to 1\,G, scanned with a high resolution on the order of 1\,mG. The ultradense loss spectrum revealed a stunning complexity with 44 resolved loss features, some of them showing up in groups and other ones appearing as isolated individual features. We also observed low-loss plateaus, which are typically a few 10\,mG wide and which are free of resonances. Here very low three-body losses facilitate highly efficient evaporative cooling.

We have studied selected resonance features in more detail by measuring the three-body recombination rate coefficient $K_3$ upon variation of the magnetic resonance detuning and the temperature. In general, the observed behavior shows strong similarities with recent observations on $p$-wave Feshbach resonances  \cite{Yoshida2018SLF,Waseem2018ULB}. At higher temperatures (above a few $\mu$K) we observed the unitarity limitation of resonant three-body losses. At low temperatures in the nanokelvin range, we observed a strong suppression of losses with decreasing temperature, provided a small detuning of just a few mG is applied. Right on top of the resonance, however, three-body losses remain very strong even at the lowest temperatures we can realize. 

Our work shows that in experiments employing fermionic $^{161}$Dy gases special attention must be payed to choosing and controlling the magnetic field in a way to avoid detrimental effects of three-body recombination losses. For our specific applications targeting at strongly interacting fermion mixtures of $^{161}$Dy and $^{40}$K \cite{Ravensbergen2020rif}, those magnetic-field regions are of particular interest where one can combine near-resonant interspecies Dy-K interaction with low-loss regions of Dy.
\begin{acknowledgments}
We acknowledge support by the Austrian Science Fund (FWF) within Projects No.\ P32153-N36 and P34104-N, and within the Doktoratskolleg ALM (W1259-N27). We further acknowledge a Marie Sklodowska Curie fellowship awarded to J.H.H.\ by the European Union (project SIMIS, Grant Agreement No.\ 894429). We thank the members of the ultracold atom groups in Innsbruck for many stimulating discussions and for sharing technological know how.
\end{acknowledgments}
\appendix
\section{\label{App:Initialconditions}Summary of the initial experimental conditions}
In Table~\ref{Table:initialconditions}, we report the experimental conditions under which the measurements reported in Figs.~\ref{fig:LowFieldRegionFBScan},~\ref{FigExampleMeasurement},~\ref{FigThreeResonancesComparison} and~\ref{FigSMFieldNoiseBroadening} have been carried out. For the measurement in Fig.~\ref{FigTemperatureDependence}, where the value of $K_3$ as a function of the $\tilde{T}$ is reported, the different temperatures have been achieved by interrupting the evaporation in a controlled way. This unavoidably has led to initial experimental conditions which vary over a wide range. The initial atom number ranges from \SI{5e4}{} to \SI{1e6}{}. The coldest samples have $T/T_F\simeq0.18$ and peak densities $\hat{n}\simeq~\SI[per-mode=symbol]{1e14}{\cm\tothe{-3}}$.
\begin{table*}[ht]
\begin{ruledtabular}
\caption{\label{Table:initialconditions}
Experimental parameters for the different sets of measurements: geometrically averaged trap frequency $\bar{\omega}$, atom number $N$, temperature $T$, Fermi temperature $T_F$, reduced temperature $T/T_F$, the function $\beta(x)$ according to Eq.~(\ref{eqBetaFactor}), and the peak density $\hat{n}$. All quantities represent the initial values after preparation (at $t=0$).}
\begin{tabular}{cccccccc}
\textrm{Figure}&
\textrm{$\bar{\omega}/2\pi\,(\SI{}{\hertz})$}&
\textrm{$N$}&
\textrm{$T~(\SI{}{\nano\kelvin})$}&
\textrm{$T_F~(\SI{}{\nano\kelvin})$}&
\textrm{$T/T_F$}&
\textrm{$\beta(T/T_F)$}&
\textrm{$\hat{n}~(\SI[per-mode=symbol]{}{\cm\tothe{-3}})$}\\
\colrule
Fig.\,\ref{fig:LowFieldRegionFBScan} & 100 & \SI{5e4}{} & 70 & 320 & 0.21 & 0.616 & \SI{4.8e13}{}\\
Fig.\,\ref{FigExampleMeasurement}(a) & 58 & \SI{1.22(4)e5}{} & 29(3) & 250 & 0.12 & 0.84 & \SI{3.1e13}{}\\
Fig.\,\ref{FigExampleMeasurement}(b) & 120 & \SI{4.3(1)e4}{} & 83(3) & 367 & 0.23 & 0.59 & \SI{5.7e13}{}\\
Fig.\,\ref{FigExampleMeasurement}(c) & 380 & \SI{1.53(3)e5}{} & 358(8) & \SI{1.77e3}{} & 0.20 & 0.63 & \SI{6e14}{}\\
Fig.\,\ref{FigThreeResonancesComparison}(a) & 88 & $\simeq$ \SI{1.2e5}{} & 64 & 379 & 0.17 & 0.72 & \SI{6e13}{}\\
Fig.\,\ref{FigThreeResonancesComparison}(b)  & 118 & $\simeq$ \SI{4e4}{} & 77 & 352 & 0.22 & 0.59 & \SI{5e13}{}\\
Fig.\,\ref{FigThreeResonancesComparison}(c) & 118 & 4~-~\SI{6e4}{} & 77~-~88 & 352 ~-~ 403 & 0.22 & 0.59 & 5~-~\SI{6e13}{}\\
Fig.\,\ref{FigThreeResonancesComparison}(d) & 118 & $\simeq$ \SI{5e4}{} & 106 & 379 & 0.28 & 0.46 & \SI{5.2e13}{}\\
Fig.\,\ref{FigSMFieldNoiseBroadening}(a) green stars & 118 & $\simeq$ \SI{4e4}{} & 77 & 352 & 0.22 & 0.59 & \SI{5e13}{}\\
Fig.\,\ref{FigSMFieldNoiseBroadening}(a,b) blue diamonds & 157 & 4.5~-~\SI{6e4}{} & 86~-~126 & 487~-~536 & 0.16~-~0.26 & 0.50~-~0.74 & 0.78~-~\SI{1e14}{}\\
Fig.\,\ref{FigSMFieldNoiseBroadening}(b) red squares & 157 & 4~-~\SI{5.5e4}{} & 88~-~187 & 468~-~520 & 0.17~-~0.4 & 0.26~-~0.71 & 0.6~-~\SI{1e14}{}\\
\end{tabular}
\end{ruledtabular}

\end{table*}
\section{\label{App:K3derivation}Extraction of the loss-rate coefficient}
Here we summarize our method to extract values for the three-body rate coefficient $K_3$ from the decay curves. Basically the same procedures have been applied in Ref.\,\cite{Ravensbergen2020rif}.
\par The particles that are more likely to collide and leave the trap are the ones in the center of the trap, with highest density and lowest potential energy. Therefore losses are accompanied by heating of the cloud, which is known as antievaporation heating in the thermal case \cite{TBRWeber2003} or hole heating in the case of a degenerate Fermi gas \cite{DFG2001Timmermans}. As a consequence the shape of the density distribution $n(\bold{r},t)$ changes with time. 
Taking a time-dependent temperature $T(t)$ into account, Eq.\,(\ref{eqAtomNumberRateEquation}) leads to a set of coupled differential equations (see for instance \cite{TBRWeber2003}). We circumvent this complication by focusing on the initial decay rate $1/\tau=-\dot N(0)/N(0)$, where $N(0)=N_0$ and $\dot N(0)$ represent the atomic number and its time derivative, respectively, both at $t=0$. For the initial decay and thus the decay time $\tau$ only the initial number density distribution $n_0(\bold{r})=n(\bold{r},t=0)$ is relevant.
\par Neglecting one-body losses and considering only the initial part of the decay, Eq.(\ref{eqAtomNumberRateEquation}) leads to
\begin{equation}\label{eqK3FromRateEquation}
    K_3=\frac{N(0)}{3\tau}\left(\int d^3r\, n_0^3(\bold{r})\right)^{-1}.
\end{equation}
For the limit of a thermal (Gaussian) distribution \cite{TBRWeber2003}, the integration results in
\begin{equation}
K_3=\sqrt{3}\frac{T_0^3}{\tau N_0^2}\left(\frac{2\pi k_B}{m\bar{\omega}^2}\right)^3. \label{eqK3ForThermalCloud}
\end{equation}
Here $\hbar$ is the reduced Planck constant, $k_B$ is the Boltzmann constant, $m$ is the atomic mass, and $T_0$ is the initial temperature of the sample.
For the number density distribution of a degenerate Fermi gas, we find
\begin{equation}
    K_3=\frac{3\pi^4}{4}\frac{1}{\tau N_0}\left(\frac{\hbar}{m\bar{\omega}}\right)^3\frac{1}{\beta(T_0/T_F)}. \label{eqK3ForDegenerateCloud}
\end{equation}
The function $\beta(T/T_F)$ is defined as the three-body integral of a finite-temperature Fermi gas normalized to the zero-temperature case:
\begin{equation}\label{eqBetaFactor}
    \beta(T/T_F)=\frac{\int d^3\bold{r}\,n^3(\bold{r})}{\int d^3\bold{r}\,n^3_{\textrm{TF}}(\bold{r})},
\end{equation}
where $n(\bold{r})$ describes the density profiles of non-interacting fermion systems at finite temperature, and $n_{\textrm{TF}}(\bold{r})$ refers to the Thomas-Fermi profile at $T=0$. By numerical integration, we find that the function can be well approximated numerically for $x=T/T_F\lesssim 1$ by $\beta\left(x\right)\simeq (1+12.75 x^2+31.05x^4-8.46x^6)^{-1}$.
\par In order to obtain the initial decay rate, we fit the decay curve with
\begin{equation}\label{eqNBodyFit}
    N(t)=\frac{N_0}{\sqrt[\alpha-1]{1+(\alpha-1)t/\tau}},
\end{equation}
which is the solution of the differential equation $\dot N/N_0=-\tau^{-1}\,(N/N_0)^{\alpha}$ for decay by few-body process of order $\alpha$. This heuristic model allows us to access the initial time decay $\tau$ without making an assumption on the true order of the loss process. The fit parameter $\alpha$ absorbs the order $n$ of the recombination process together with effects from heating. The initial atom number $N_0$ is also derived from the fit, whereas the initial temperature $T_0$ is measured separately. The value of $K_3$ is finally obtained from Eq.~(\ref{eqK3ForThermalCloud}) or Eq.~(\ref{eqK3ForDegenerateCloud}).

\section{\label{App:Broadening}Broadening Effects}
\begin{figure*}[t]
\centering
\includegraphics[scale=1.6, width=2\columnwidth]{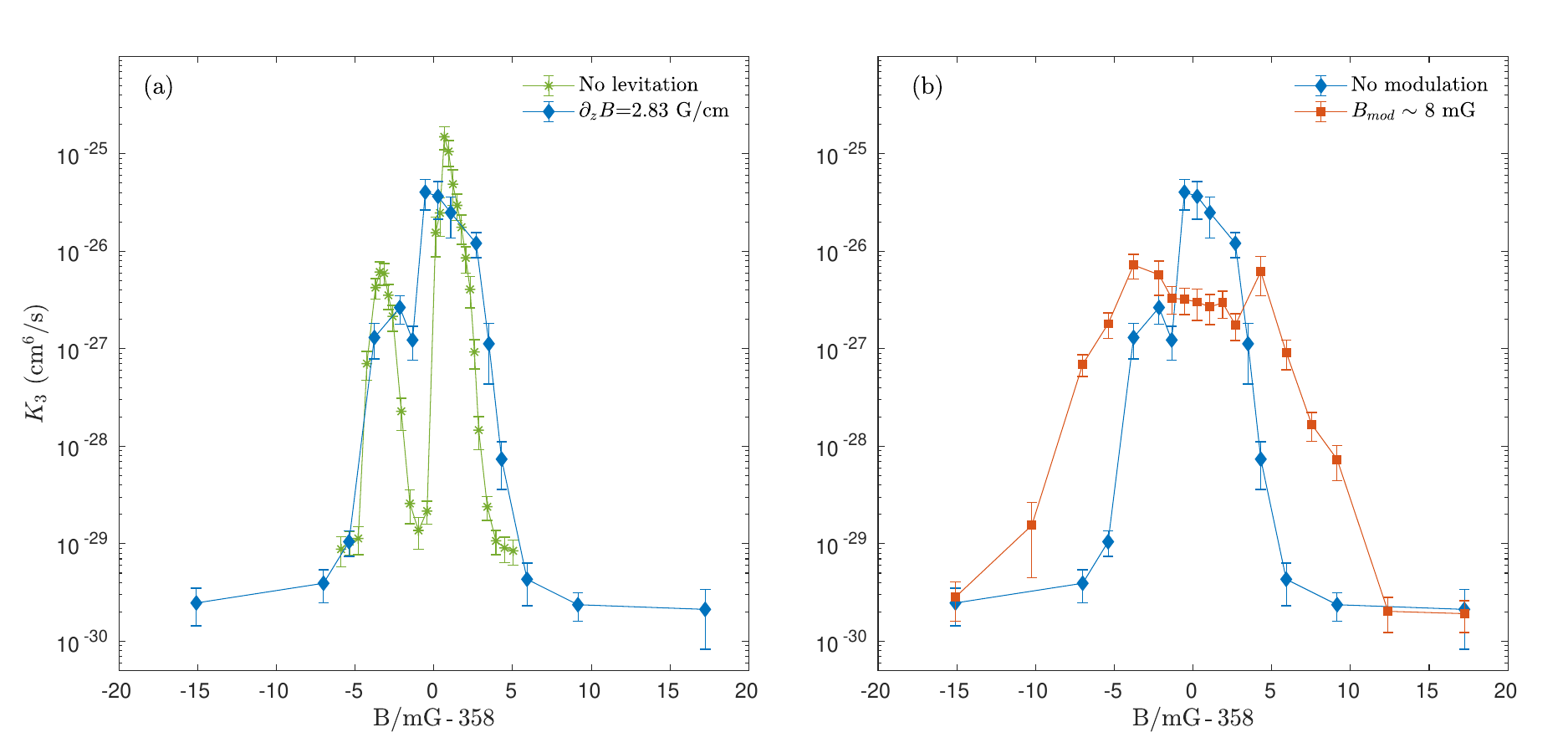}
%\captionsetup{justification=justified}%singlelinecheck=false}
\caption{\label{FigSMFieldNoiseBroadening} Effect of magnetic field broadening. The $K_3$ value is plotted versus the magnetic field, in the region around \SI{358}{\milli\gauss}. (a) Effect of the magnetic field gradient used for levitating Dy. (b) Effect of a sinusoidal magnetic field modulation, with a frequency of \SI{600}{\hertz} and an peak-to-peak amplitude of 8~mG. } \end{figure*}
When dealing with a Feshbach spectrum dense of narrow resonances, it is important to understand and possibly eliminate potential broadening effects. In our system, we identify two sources of broadening: magnetic levitation and magnetic field noise. In experiments where a decrease of the trapping frequencies leads to a trapping potential not deep enough to hold atoms against gravity, magnetic field levitation is often used to cancel (or reduce) the gravitational sag \cite{Weber2003BEC}. However, the presence of a magnetic field gradient introduces an inhomogeneity of the magnetic field along the vertical extent of the cloud. Assuming full levitation for dysprosium ($\partial_z B=\SI[per-mode=symbol]{2.83}{\gauss\per\centi\meter}$) and a typical Thomas-Fermi radius $\sim\!\!\SI{10}{\micro\meter}$, our atomic sample is subjected to a magnetic-field variation of $\sim\!\SI{6}{\milli\gauss}$ over the trap volume. In Fig.\,\ref{FigSMFieldNoiseBroadening}(a) we demonstrate the effect of levitation broadening on the $K_3$ coefficient. We work at the 358-mG resonance. If no gradient is applied we can resolve a double-peak structure. The two features have a full width of about \SI{2}{\milli\gauss}, with peak values of \SI[per-mode=symbol]{6(3)e-27} and \SI[per-mode=symbol]{1.5(4)e-25}{\cm\tothe{6}\per\second}, respectively. The presence of the magnetic field gradient reduces the resolution and the two peaks almost merge into a single feature 7.7-mG wide. We can still distinguish two local maxima, whose values are reduced with respect to the no-levitation case.
In view of this broadening effect, we decided to carry out all measurements reported in the main text without magnetic levitation. To achieve low enough trap frequencies in the shallow trap where the atoms are transferred at the end of the evaporation, we employ the second ODT stage with an increased waist of the horizontal trapping beam, as mentioned in Sec.\,\ref{Subsec:SamplePreparation}. Such a trap geometry allows us to reach low trapping frequencies without too much sacrificing the trap depth.
\par In another set of measurements, we investigate the effect of magnetic field fluctuations. We artificially introduce noise into the system by adding a sinusoidal magnetic field modulation to the bias field. We chose a modulation frequency of $\SI{600}{\hertz}$, i.e.~faster than the trap frequencies, but still slow enough to avoid technical complications. We then measure decay curves for different modulation strengths and magnetic field detunings. In Fig.\,\ref{FigSMFieldNoiseBroadening}(b) we report the obtained $K_3$ values versus magnetic field detunings for 10-mG peak to peak modulation. As a reference, we plot the $K_3$ profile measured in the absence of artificial noise. 
The modulation results in a broadening of the feature and consequent loss of field resolution. The peak value decreases and the curve flattens off.
\par Regardless of the source of broadening, a magnetic-field inhomogeneity (in time and space) has an averaging effect on the $K_3$ coefficient, and leads to a broadening and weakening of the narrow loss features characterizing the dysprosium Feshbach spectrum.

\bibliography{referenceFromThesis, references, ultracold, rudi_newrefs}

\end{document}